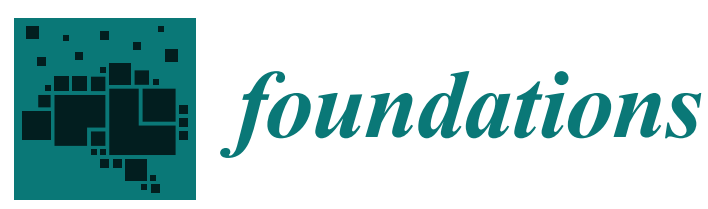

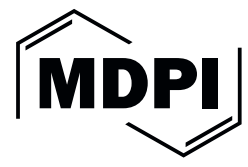

*Article*

# Flat Bundles on Function Manifolds and Evolution Equations in Quantum Field Theories

**Stanislav Srednyak**

Department of Physics, Duke University, Durham, NC 27708, USA; stan.sredn@gmail.com

**Abstract**

In this paper, we discuss extensions of the canonical quantization procedure in quantum field theories. We focus specifically on S-matrix representation as a T-exponent. This extension involves flat bundles on certain infinite dimensional functional manifolds of local time. The motivating problem is first principles treatment of bound states in quantum chromodynamics as well as precision physics of the hydrogen atom and the muonium. Our main results include systematic treatment of flat bundles in an infinite dimensional setting, generalization of Hamiltonian evolution and functional renormalization group evolution equations in quantum field theories. We discuss several results from finite dimensional theory that have analogies in the functional setting. This includes construction of moduli space of flat connections and isomonodromic deformations. One of the outcomes of our analysis is a construction of a rich family of functional flat bundles with rational connections. This class of connections exhibits a rich set of mathematical properties. In particular, we construct examples of the fundamental groups of spaces which have a definable continuum of generators. Physical states correspond to points in the moduli space of bundles on these spaces. On the physics side of things, we conclude that spacetime notions, such as spaces of particle configurations, emerge effectively as spectral sets of functional differential operators.

## 1. Introduction

Perturbative quantization of quantum field theories is a well defined procedure. It gives an algorithmically defined set of functions; while these functions are often hard to compute and lead to a variety of interesting mathematical problems, they are well defined. Problems start to arise when one tries to extend these ideas to nonperturbative objects, such as bound states in QCD, or even in QED.

In this paper, we discuss an approach to this problem based on functional evolution equations. The idea of local time was behind the canonical quantization approach from the very beginning of quantum field theory. In the approximation used in these papers, the calculations reduce to the perturbative S-matrix. However, this approximation does not really work in strongly coupled confining systems such as QCD. Therefore, it becomes justified to have a more detailed look on possible evolution equations with local time. This is what is done in this paper.

One other important departure from the classical viewpoint is consideration of more general forms for the second quantized fields. Canonical quantization conditions suggest that creation/annihilation operators can be interpreted as functional derivatives with respect to a functional argument, denoted as $a(\vec{p})$, for the electron operators. This interpretation brings into consideration the manifold of functions $a(\vec{p})$. In the usual perturbative treatment, this function manifold is just the affine space of all functions. For example, they can be square integrable. This manifold does not play any role in the final results of perturbative calculations. It completely drops out of the consideration after we take matrix elements of the S-matrix between perturbative asymptotic states, and then by using the commutation conditions, completely eliminate all mention of the function variables. This is not the case in our approach. We systematically study this space and the first order evolution equations on it. It should be mentioned that there are also higher order evolution equations possible, but we postponed their consideration to future publications. Here we only mention that first order equations are equivalent to flat bundles on functional manifolds, and thus are topological, while higher order equations are more geometric and as such require more sophisticated theory.

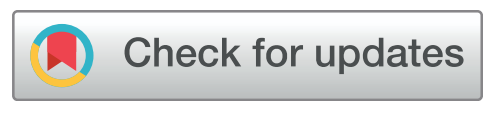

We summarize the physics contributions of this paper:

(1) We introduce a novel interpretation of field operators as functional differentiation operators. This interpretation allows us to develop mathematical formalism for analysis of particle scattering processes. This allows us to relate QFT (notably, QED) to functional differential operators. This opens the way to develop and use functional analogies of pseudodifferential operators.

(1.1) We introduce and rigorously discuss the notion of functional manifold. Our developments suggest that conventional perturbation theory saw just a small flat coordinate patch of this manifold. The nonperturbative version of the theory, as discussed in our paper, is sensitive to the topology and global structure of this manifold.

(1.2) We introduce, and demonstrate the physical necessity, of the internal 3-dimensional manifold $M^3_{\vec{p}}$, from which the above functional manifold is constructed. This manifold can be thought of as a nonperturbative analogy of the complexification of the usual manifold of on-shell 3-momenta. But in our formalism it is a more flexible object. We demonstrate how its properties can be inferred from observable data.

(2) We show that the usual S-matrix of QFT can be generalized into an invariant of a flat bundle on the above functional manifold $\mathbb{M}$. Scattering amplitudes are obtained as expansion into the basis of integrals over closed loops inside the functional manifold of the flat connection that defines the theory. Since flat connections are rigid, this explains why physics observables are essentially parameter free.

(3) Our mathematical formalism naturally realizes the idea that physical spacetime is "emergent". This is an old idea that has been discussed a number of times [1]. In our formalism it occurs naturally. Spacetime "emerges" when particle dynamics are sufficiently simple and can be approximated by factorizable functionals.

(4) Our treatment is nonperturbative. It can be applied to bound state problems in QCD, which was indeed the main motivation for developing this formalism. We take first steps in this direction. In particular, we discuss how we can sum infinite towers of correlated gluon wave functions that contribute to the mass of the proton. Rigidity of flat connections is an important feature for fixing the equations for the towers of partonic wave functions.

We now summarize the mathematical contributions of this paper:

(1) We defined and discussed the notion of infinite dimensional manifolds that play a role in QFT. We paid special attention to spaces with variable smoothness.

(2) We defined flat bundles on functional manifolds and demonstrated that they exist in a wide range of function manifolds.
(3) We defined the notion of moduli of flat bundles on function manifolds and a class of invariants of these bundles. We demonstrated that this class plays a special role in physics by connecting it to amplitudes. In the Appendix A, we drew a parallel to non-Abelian cohomology and rational homotopy theory. We demonstrated that the evolution equation in non-Abelian cohomology is well defined in the infinite dimensional setting. On an example of manifolds modeled on sequence spaces, we demonstrated the use of descriptive set theory for the formulation of results concerning the structure of flat bundles.
(4) In the Appendix B, we studied an example of a manifold of resurgent functions with regular singularities. We proved a result that connects the algebra of singularities of the flat connection and the functions in the manifold to the algebra of singularities in the invariants of the connection. This result provides a nonperturbative version of the singularity structure in perturbative amplitudes.
(5) We introduced the notion of effective spacetime operators, that act on the sections of flat bundles and take values in the category of finite dimensional manifolds. We showed that modern QED spectroscopy experiments should be interpreted using this notion of effective spacetime operators. We showed how the notion of spacetime, as used in these experiments, arises as a certain approximation from the infinite dimensional bundles on which the quantum dynamics take place. This notion of effective spacetime operators has a broader mathematical relevance, as it also plays a role in the studies of topology and geometry of infinite dimensional manifolds, playing the role of dimension, degree, topological index, and other discrete characteristics.

## 2. Physical Preliminaries

The applications of this paper include hadronic spectroscopy and precision physics of bound states in QED. We consider the latter to be a very important testing ground. We would like to start by mentioning that there is a mature framework for bound state computations, see, e.g., [2–6].

It is known that perturbative QED faces challenges when extended beyond two loops (specifically, applied to bound state problems, in contrast to single electron anomalous momentum calculation, which involves completely different physics, both from experimental and from theoretical points of view), particularly in consistently incorporating contributions from multiparticle virtual states. Our goal is to investigate a geometric-functional framework where such contributions can be systematically treated through the structure of wave functionals and flat connections. Inclusion of such objects implies consideration of wave functions of the type $\psi_{\alpha\beta\gamma}(x, y, z)$. There are problems with formulating gauge invariance of such objects. In particular, full formalism requires consideration of mixed type wave functions $\psi_{\alpha\mu}(x, y)$. This is related to the known multi-body Dirac Equations [7], and is an open problem.

The $g - 2$ value of the electron as calculated in QED [8] agrees well with observation [9]. This is not so for the muon [10]. There is widespread understanding that hadronic QCD corrections play a role. Studies were mostly confined to phenomenological and lattice models [10]. It is our goal to set foundations for analytic study. Such foundations in particular requires understanding mechanisms of confinement. We will show that the rigorous treatment of this problem involves certain types of resurgent functions akin to the ones considered in holomorphic dynamics [11,12]. These functions arise as solutions to functional equations we consider below.

For the electron, while g − 2 treatment is experimentally valid at tenth order, the agreement of the Lamb shift is only tested [13] at one-loop accuracy (we note here that one loop contribution includes many orders in $\alpha$ [14]). At the moment it is conceivable that there are experimental deviations at two-loop accuracy, with the conventional model of the proton radius terms [15]. The Lamb shift for muonium exhibits similar phenomena [16].

Development of analytic models for the multiparticle bound states includes the major unsolved problem in atomic physics, namely, analytic methods for the atomic spectra in quantum mechanics. The current literature on this problem [17,18] includes various approximate models such as coupled cluster models, and active space methods. None of these is suitable for our approach. What we really need is quantum KAM theory in the sense of [19] together with function analytic methods [20,21]. The development of analytic tools for bound states may eventually require techniques from quantum KAM theory. However, in the present paper we restrict ourselves to foundational geometric considerations and postpone application of full-fledged quantum KAM formalism to future studies. Although it is impossible to treat quantum KAM in any detail in this paper, we mention that it is a theory to treat non-linear functional evolution PDEs of the form $\frac{dp(x)}{dt} = \frac{\delta H[p,q]}{\delta q(x)}, \frac{dq(x)}{dt} = -\frac{\delta H[p,q]}{\delta p(x)}$. This formalism is applicable to non-linear multiparticle equations of Schrodinger type $id_t\phi(t, x_1, \ldots, x_n) = \Delta_{x_1,\ldots,x_n}\phi + \lambda\phi^2 + O(\phi^3)$. The formalism does not address the linear problem, which must be treated by methods of spectral theory.

There exist at least two functional approaches to bound state problems. Both have roots in functional evolution equation

$$i\partial_t F = \hat{\hat{H}}(t)F\,. \tag{1}$$

One of them is explored in this paper. The other approach leads to systematic study of functional PDEs of the form

$$\hat{\hat{L}}F = 0 \tag{2}$$

where $\hat{\hat{L}} = \hat{\hat{L}}[a, \frac{\delta}{\delta a}]$ is a functional differential operator. The functional $F$ should be interpreted as a bundle on a functional manifold. These equations are of more geometric nature. For example, with a suitable choice of $\hat{\hat{L}}$ they can represent functional analogies of holonomic D-modules, and therefore generalize the flat connection equations we consider. If the operator $\hat{\hat{L}}$ is scalar, the solutions depend on boundary conditions. This approach is considerably more demanding in terms of mathematical apparatus. In particular, full treatment requires development of an elliptic package on a functional case. In this way, it contrasts the approach we consider in this paper. The notion of flat bundles is sensitive only to the topological structure, and is simpler for analysis. The approach of Brodsky et al. [22] is in fact within this functional PDE's framework. What we show in this paper is that with certain choice of function manifold, it is possible to reproduce the calculations of $g-2$ to high order and of Lamb shift to leading Bethe log approximation.

Algebraic QFT [23,24] is a very rich mathematical theory. It has a number of interplays with the theory laid out here. First, it explicitly discusses the role of fields of C* algebras. We extend this notion by proposing to consider fields of integral operators with holomorphic kernels. This is phenomenologically well motivated, as we explain in later sections. Second, this theory is probably one of the first theories to consider abstract combinatorial spacetimes (e.g., Section 6 in [1,23]), or emergent spacetimes, in our terminology.

Functional Renormalization Group [25,26] has been applied to QCD. In ref. [27] it was applied to the bound state problem. Our approach is an extension of this in the sense that the taylor coefficients of the solutions of the flat bundle equation are more complicated than the fixed complexity ansatzes considered in these references. The idea of Functional RG is very naturally inherent to QCD bound state theories. These bound states comprise multiple energy scales, and correspondingly there are couplings at many scales. Functional RG allows us to fix these scales self consistently, normalized to the energy density in the bound state.

## 3. Functional Manifolds and Functional Differential Operators

We will discuss certain open functional spaces and differential operators on them. In order to set the stage for this discussion we need some more general discussion of function spaces. In finite dimensions, there is a single invariant that characterizes the space—its dimension. This is not so in infinite dimensions. There is a zoo of spaces in the precise mathematical meaning that the set of Banach subspaces of $C([0,1])$ forms an analytic family [28]. This set is unclassifiable in ZFC. Note that the spaces in this family are realizable as spaces of sequences. In the context of spaces of functions on a differentiable manifold, the corresponding phenomenon is discussed in [29]. Intuitively, one can think of spaces of functions with varying degrees of differentiability along closed subsets of the underlying finite dimensional manifold.

The above phenomenon has an immediate bearance on physics. The path integral $\int_{C^L(M)} exp(S_{eff}[f])Df$ depends on the space of integration $C^L(M)$ where $L$ denotes one of the above Frechet structures. This dependence is manifest if the effective action functional is sufficiently complex, for example, a non-local integral polynomial with nondegenerate measurable coefficients.

This phenomenon of dependence on the function space of integration is not manifest in the usual QFT phenomenology. In particular, it is never part of the discussion in QED calculations or QCD calculations. We propose an explanation for this in the following. The functionals studied in the literature, namely, QED or Yang–Mills functionals, are very specific. The functions that result from perturbation theory have essentially canonical functional form. They are locally represented by convergent logarithmic Puiseux series (this is true for arbitrary Feynman diagrams). Therefore, what has been dealt so far in physics is a definable class of Banach spaces that are given by convergent Puiseux series on a finitely definable family of submanifolds of the configuration space.

We will use this definition in the following and consider function manifolds modeled on spaces of convergent logarithmic Puiseux series on open charts. A consequence of this choice is that the dual space also has this structure. We will be dealing with reflexive Banach spaces. It is known that the tensor product on Banach spaces is not unique [30]. In the case of the Hilbert space, there is a connection to the Connes embedding problem [31]. However, in our case, the tensor product is unique. The tensor product of two functions each represented by a Puiseux series on a complement to a set of varieties is again a function represented by log-Puiseux series on the product manifold. Precise formulation of these results as well as their proofs is somewhat long and technical, and we have to postpone its discussion to future publications.

For doing analysis on manifolds, the so-called elliptic package is essential. This elliptic package includes the ability to define Green's functions and, as a prerequisite, the ability to integrate over the manifold. In our infinite dimensional setting we therefore are hitting the famous problem of defining measures on infinite dimensional spaces. Rigorous treatment of measures on infinite dimensional manifolds remains an open problem in this context. We restrict our current analysis to topological aspects of flat bundles, for which the notion of measure is not essential. Nevertheless, further development of this theory will require precise integration theory on functional spaces, which we plan to address in forthcoming work. For this theory, we only need the following ingredients:

(1) Intersection theory of finite and countably infinite codimension 1 holomorphic subvarieties.
(2) Integration of closed twisted forms along lines on the infinite dimensional manifold.

This theory is substantially simpler than the theory of functional differential equations, and it does not involve functional measures. The theory is essentially topological. The topology here refers to the topology of infinite dimensional spaces. It should be noted that many infinite dimensional manifolds are contractible, e.g., spheres [32], and general linear groups [33]. This is not the case for the spaces we consider.

## 4. Flat Bundles on Functional Manifolds

In this section, we lay the physical foundations of functional evolution equations.

### 4.1. Functional Evolution Equations

In this section, we recall how the conventional S-matrix is obtained from functional flat bundles. The functional Schrodinger equation

$$i\frac{\partial}{\partial t}F(t) = H(t)F \tag{3}$$

leads to the scattering equation

$$F(+\infty) = SF(-\infty) \tag{4}$$

where $S = Texp(\int H(t)dt)$ is the scattering matrix. This equation has a formulation with local functional time [34–36]

$$i\frac{\delta}{\delta T(\vec{x})}F = H[T]F \tag{5}$$

where the local time $T(\vec{x})$ can be identified with a Cauchy surface restricted by causality condition [34,37,38].

### 4.2. Choice of Functional Variables for Functional Evolution Equations

There are several versions of flat bundle equations in functional settings, of increased complexity. We consider them in turn. The simplest version is

$$\frac{\delta}{\delta T(q)}F_i[T] = \sum_j H_{ij}(q)[T]F_j[T]\,. \tag{6}$$

This bundle is finite dimensional. The next in simplicity example is

$$\frac{\delta}{\delta T(q)}F(p)[T] = \int (dr)H(q;p,r)[T]F(r)[T]\,. \tag{7}$$

Of most physical relevance is the next case

$$\frac{\delta}{\delta T(q)}F(p_1,\ldots,p_n)[T] = \sum_m \int (dr_1\ldots dr_m)\times \tag{8}$$

$$\times H(p_1,\ldots,p_n;r_1,\ldots,r_m)[T]F(r_1,\ldots,r_m)[T]\,. \tag{9}$$

We can write it in condensed notation as

$$\frac{\delta}{\delta T(q)}F[T,a] = \Omega(q)[a,\frac{\delta}{\delta a}]F[T,a] \tag{10}$$

where $a = a(p)$ runs over a suitable function space of the Puiseux series. In general, the variable $p$ of $a(q)$ can run over a different finite dimensional manifold than $q$, the variable of $T(q)$. This observation will be essential later in our treatment of emerging spacetime. The space of variables $p$ in the functional equation does not necessarily coincide with the physical momentum space. The physical momentum space is a linearization of the space of abstract momenta $p$ on which the functional variables are defined.

### *4.3. Definition of a Functional Flat Bundle*

Consider the functional bundle on functional manifold $\mathbb{M}$. By the functional flat bundle we mean an object glued from local trivializations with the structure group $C^{\mathbb{L}_1}(M_1) \otimes_\tau C^{\mathbb{L}_2}(M_1)$. The manifold $M_1$ can in principle be (and often is) different from the manifold $M$ underlying the function manifold $\mathbb{M}$. For example, in physics applications often $M_1 = M^n$—this corresponds to the configuration space of $n$ particles. The symbols $\mathbb{L}_i$ denote the smoothness class of functions we would like to consider. In geometric situations [39,40] these are often $C^{1,\alpha}$, for example, Lipschitz classes. In physics applications, these are often $C^\infty_{log}(M; D)$—functions that are holomorphic away from a tame divisor $D$, along which they can have logarithmic singularities (we explicitly allow infinite dimensional bundles with structure groups, some geometrically defined $C^*$-algebra, e.g., group algebra of $\pi_1(M - D)$, with a countable set of generators). $\mathbb{L}_1$ is often dual to $\mathbb{L}_2$, e.g., it can denote Bounded Variation (BV) if $\mathbb{L}_2 = 0$. $\tau$ is one of the tensor products [30]. In applications to QCD, this could be one of the mold tensor products of spaces of resurgent functions with logarithmic singularities [41].

Let the local trivialization of the bundle be given by the functionals $F_\xi[f] \in C^{\mathbb{L}_2}(M_1)$. Then for this trivialization, the flat bundle takes the form

$$\frac{\delta}{\delta f_x} F_\xi[f] = \int d\eta \Omega_{\xi,\eta}(x)[f] F_\eta[f] \,. \tag{11}$$

We now make a comment on the smoothness class of the functional $\Omega[f]$. For topological applications, it makes sense to consider the $C^\infty(\mathbb{M})$ class. This is especially interesting for iterated diffeomorphism spaces $Diff(Diff(M))$, the spaces of functional diffeomorphisms of the usual diffeomorphism spaces $Diff^{\mathbb{L}}(M)$, with smoothness $\mathbb{L}$. But in physics applications, the connection usually can have logarithmic singularity along a functional divisor with components $P_a[f]$. In local coordinates these connections look like

$$\Omega_{\xi,\eta}(x)[f] = \sum_a \frac{A_{a,\xi,\eta}[f]}{B_a[f]} \tag{12}$$

for some integral polynomials $B_a[f]$.

### *4.4. Note on Variable Smoothness*

Most of our discussion applies to the case that the functional manifold $\mathbb{M}$ has variable smoothness, i.e., the Banach type of the tangent space changes from point to point. This type of example emerges naturally in physics applications: the tangent space has type $C^{\mathbb{L}[f]}(M)$ where the smoothness type $\mathbb{L}[f]$ is in fact varying with the point $f$. Examples of this type are ubiquitous in applications to variational problems in geometry. For example, it is easy to get functions of type $C^{\alpha(x)}(M; C)$ where $C$ is a tamely embedded Cantor set in $M$ and $\alpha(x)$ is a point dependent Holder exponent. This example also can be extended to holomorphic settings, using resurgent functions, even when $dim_{\mathbb{C}}(M) = 1$. The examples mentioned above indicate that Frechet structures in the underlying manifold $\mathbb{M}$ and the bundle may be incompatible. Our results still hold in this case.

### 4.5. Classification Theorem and Particle Interpretation

One of the main theorems in the study of flat bundles in finite dimensional case is the classification of the bundles in terms of representation theory of the fundamental group [42]. The bundle is classified by the representations

$$\rho : \pi_1(M) \to G \tag{13}$$

where $G$ is the structure group.

The aim of this paper is to pose the following hypothesis: there exists a functional manifold $\mathbb{M}$ and a structure group $\mathbb{G}$, such that bound states correspond to representations of the fundamental group $\pi_1(\mathbb{M})$ in $\mathbb{G}$.

We do not prove any analogy of this theorem in our situation. Full discussion would entail discussion of lattices in Frechet algebras. Our aim is to provide evidence in favor of this hypothesis, by demonstrating that existing calculations fall within this framework. Furthermore, we analyze a few purely mathematical situations where such constructions are fruitful.

### 4.6. Quantization Surfaces

Consider Cauchy surfaces $\sigma_\mu(\xi)$ in the complex spacetime $\mathbb{C}^d$. We consider the following differential operators

$$L(\zeta)[\sigma, a(p), \frac{\delta}{\delta a(p)}] = \sum \int_{\sigma_1} \dots \int_{\sigma_n} L(\zeta; \xi_1, \dots; \eta_1, \dots) a(\xi_1) \dots \frac{\delta}{\delta a(\eta_1)} \dots . \tag{14}$$

The evolution equation

$$\frac{\delta}{\delta \sigma_\mu(\zeta)} F[\sigma, a] = L_\mu(\zeta)[\sigma, a(p), \frac{\delta}{\delta a(p)}] F[\sigma, a] \tag{15}$$

The space of Cauchy surfaces on Lorents manifold has an ideal boundary composed of submanifolds that have light-like elements. In the case when the underlying manifold $M$ is the usual affine Minkowski space, this geometry is rather simple: the space of cauchy surfaces is contractible. It is the convex geometry of its ideal boundary that is of physics interest; we expect to extend the bundle to functional Fulton–MacPherson (fFM) compactification of this space, and we expect logarithmic singularities along the components of the compactification divisor. In the case of topologically nontrivial Lorenz $M$, the space $Cauchy(M)$ can have many components. Low codimension strata of fFM space are given by light cones with multiple vertices.

## 5. Moduli Spaces of Flat Bundles

In this section, we discuss the construction of moduli space of flat bundles. In finite dimensions, the literature on this subject is vast [43–46]. There are results on the Poisson geometry of the moduli space [44,46]. In the case of Riemann surfaces, there is a beautiful connection with Goldman bracket [47] and string topology [48]. In the case of bundles on the line, there is extensive theory of isomonodromic deformations [49], that in particular leads to the theory of Painleve Equations [50] and irregular Riemann Hilbert problems [51]. The generalization of this theory already in two dimensions is not known. In our case, we are dealing with infinite dimensional generalization of this theory. Naturally, there are certain mathematical difficulties that we encounter. However, as we will show heuristically, these difficulties can be surmounted.

We will start with a heuristic construction of moduli space of flat bundles in finite dimensions. The equations are of the sort

$$\partial_\mu f(x) = \sum_I \frac{A_{I,\mu}}{L_I(x)} f \tag{16}$$

where $A, L$ are polynomials. In local coordinates $z_I = L_I$ the equations have the form

$$\frac{df}{dL_I} = (\frac{C_I}{z_I} + R_I(z))f \tag{17}$$

where $C_I$ are constants and $R_I$ are regular at $z = 0$. These $C_I$, constructed for each zero-dimensional intersection point of $L_I$, form a coordinate system on the space of flat bundles. This in particular demonstrates finite dimensionality of moduli space if there are no singularities at infinity. There are more sophisticated coordinate systems that can be constructed using gauge invariant integrals of the connection [52]. These systems lead to higher dimensional notions of the associator [53].

In infinite dimensions, the situation is similar. We consider as an example the finite dimensional bundle

$$\frac{\delta}{\delta T(q)} F_i[T] = \sum \frac{A_{I,i,j}(q)[T]}{B_I[T]} F_i[T] \tag{18}$$

with polynomial functionals $A, B$. Using local coordinates $z_I = B_I[T]$, the system reduces to

$$\frac{\partial}{\partial z_I} F = (\frac{C_I}{z_I} + R_I[z])F \tag{19}$$

with constant $C_I$. These constant matrices $C_I$ form a local chart on the moduli space of flat bundles. The distinction from the finite dimensional case is that in general we need to solve polynomial integral equations $B_I[T] = 0$. For generic $B_I$, there are infinitely many solutions.

### 5.1. Rational Flat Bundles

In this section, we consider that case when the evolution operator can be considered as a rational function of time $T$:

$$\Omega(q)[T, a, \frac{\delta}{\delta a}] = \sum_I \frac{A_I(q)[T, a, \frac{\delta}{\delta a}]}{B_I(q)[T, a]} \tag{20}$$

where $A, B$ both have polynomial dependence on $T$. For example, for $B$

$$B[T, a] = \sum \int B(q; q_1, \ldots, q_n)[a]T(q_1)\ldots T(q_n)(dq_1 \ldots dq_n) \tag{21}$$

The flatness condition is

$$\frac{\delta}{\delta T(q_1)}\Omega(q_2) - \frac{\delta}{\delta T(q_2)}\Omega(q_1) = [\Omega(q_1), \Omega(q_2)] \tag{22}$$

Classification of solutions to these equations depends sensitively on the intersection theory of the function manifolds $\Lambda_I = \{a : B_I[a] = 0\}$. A physically interesting situation arises when there are enough conditions to get zero dimensional intersections, i.e., there is a countable infinity of $B_I$. Linear terms of the equations $B_I[a] = 0$ are integral equations of the form $\int B_I(p)a(p) = 0$. For physics applications it is important to consider the case of holomorphic $B_I(q)$ with logarithmic branchings along submanifolds: $B_I(p) = R_1 L^b + R_2$ where $R_i$ are regular functions and $L$ are the local equations for the branching locus. Then the solutions of these equations have the same property of having logarithmic branchings along a set of locally holomorphic manifolds. The singularity locus of the solutions is determined in an algebraic way through the singularity loci of the functions $B_I(q)$. In general, it has infinitely many components. The branching exponents of the solutions are determined as linear combinations of the branching exponents of the functions $B_I$. The local analysis of logarithmic series expansions for the solutions $a(p)$ of $B_I(p)[a] = 0$ gives a practical way to obtain a discrete set of solutions.

### 5.2. Exponential Flat Bundles

In this section, we give an example in the spacetime domain in which consistency conditions can be solved. Our starting point is the equation

$$\frac{\delta}{\delta T(q)} F = H(q)[T]F \,. \tag{23}$$

Where we assume that $H(q)[T]$ has the following exponential ansatz

$$H(q)[T] = \sum H_n(q) exp(\int (dp)\omega_n(p)T(p)) \,. \tag{24}$$

Here $n$ runs over a discrete semigroup. For example, it can be $\mathbb{N}^d$. The consistency conditions are

$$\omega_n(p)H_n(q) - \omega_n(q)H_n(p) = \sum_i [H_i(p), H_{n-i}(q)] \tag{25}$$

The sum in this equation is finite. Another class of functional flat bundles with exponential singularities is given by

$$\frac{\delta}{\delta T_q} F = \sum_I \sum_{(n_i)} \frac{A_I(q)}{\prod_{i\in I} B_i^{n_i}[T]} F \tag{26}$$

where we allow $n_i > 1$.

### 5.3. Two Constructions of the Bundle

In this section, we consider exclusively the bundles defined by the equation

$$\frac{\delta}{\delta f_z} F = \Omega_z F \tag{27}$$

with $F = \oplus F_{x_1,\ldots,x_n}[f]$ and $\Omega_z = \oplus\Omega(z; x_1, \ldots, x_n; y_1, \ldots, y_m)$ of rational integral type. We suppose that the components $L_a[f]$ of singularity divisor $\Omega_z$ intersect generically at points $f_A$, $A = (a_1, \ldots, a_n, \ldots)$. In this setting, there are two ways to construct the bundle:

(1) Solve consistency conditions for $\Omega_z$ at all points $f_A$.
(2) Use the representation theoretic data $\pi_1(\mathbb{M}) \to \oplus C(M) \otimes_\tau \ldots \otimes_\tau C(M) \otimes_\tau C'(M) \otimes_\tau \ldots \otimes_\tau C'(M)$. The second construction is tightly related to the invariants that we consider in Section 5.5.

These two descriptions of the bundle must be equivalent. This defines us a map

$$Texp(\int_t \int_z \Omega_z f(z,t)) \to \oplus_A \Omega_z(x_1, \ldots; y_1, \ldots)[f_A] \tag{28}$$

that maps the monodromy data to the functional algebraic data at each of the intersection points $f_A$. This map is a highly transcendental functional map. Its function theoretic properties are key for understanding properties of bound states in our theory.

*5.4. Solutions in Terms of Series*

In this section, we would like to construct a family of solutions to the flat equations

$$\frac{\delta}{\delta T(q)} F_i[T] = \sum_I \frac{A_{I,i,j}(q)[T]}{B_I[T]} F_j[T] \,. \tag{29}$$

The solutions in question have the form

$$F[T] = \sum_{N=(n_I)} F_N \prod B_I^{\alpha_I + n_I}[T] \,. \tag{30}$$

In the following we explain the structure of this solution. This solution is valid near a zero-dimensional intersection $a_0$ of functional varieties $B_I[a] = 0$. The constants $\alpha_I$ are eigenvalues of the matrices $A_{I,i,j}[a_0]$. We assume that these matrices are simultaneously diagonalizable. The case of non-diagonalizable matrices $A_I[a_0]$ is very interesting and has been studied in finite dimension in [54]. The constants $F_N$ depend on the choice of the point $a_0$.

The constants $F_N[a_1]$ and $F_N[a_2]$ at different zero-dimensional intersections are related by monodromy transformation:

$$M = Texp \int d\tau \int (dp)\Omega(p)\delta T(p) \tag{31}$$

where the outer integration connects the points $a_1$ and $a_2$ by a line in the space $T(p)$ avoiding the singularity locus $B_I[a] = 0$.

We mention another extreme case where the theory is interesting—the case when there are just two components of the singularity locus:

$$\Omega(p) = \frac{A_1(q)[a]}{B_1[a]} + \frac{A_2(q)[a]}{B_2[a]} \tag{32}$$

Construction of solutions in this case involves classification of flat bundles on the intersection $\{B_1 = 0\} \cap \{B_2 = 0\}$.

*5.5. Invariants of Flat Bundles*

The theory of flat bundles in finite dimension comes with a set of invariants that are constructed as ordered integrals along paths that connect singularities. In particular, in the theory of KZ equations, the limit $lim \epsilon^{-A} \int_{\epsilon_1}^{1-\gamma} Texp(\Omega(t)dT)\gamma^{-B}$ plays a role. There are analogous quantities in the theory of functional flat bundles:

$$V\{\Omega\} = \prod \epsilon_I^{\alpha_I - 1}(Texp \int_{a_1}^{a_2} \int (dp)\Omega(p)\delta T(p)) \prod \gamma_J^{\alpha_J - 1} \,. \tag{33}$$

The integral is extended between two zero-dimensional intersections $a_1$,$a_2$.

*5.6. Flat Bundles on Complete Intersections*

Flat bundles with logarithmic singularities on complete intersections in Kahler manifolds are relatively well understood. They are essentially reconstructed from flat bundles on one-dimensional intersections of components of the logarithmic divisor. The latter are flat bundles on punctured Riemann surfaces of finite genus.

We consider similar procedures in infinite dimensions. Suppose that the function manifold $\mathbb{M}$ is holomorphic and that the singularity divisor of the functional bundle is also non-singular holomorphic. Then, as our local calculations above show, we can expect that the bundle is reconstructible from its pullbacks on the singularity components of the singularity locus. There is a map

$$R : (\pi_i^* F) \to F \tag{34}$$

where $\pi_i^* F$ contain first $d_i$ coefficients in the logarithmic expansion of $F$. Each $\pi_i^* F$ is a flat bundle on the singularity component $L_i$. For reconstruction, only the leading coefficient suffices.

We can iterate this process until we arrive at one-dimensional intersections $R_I = \cap_{i \in I} L_i$. In general, these 1d intersections are of wild type, as examples show. The examples can be constructed in sequence spaces, where such intersections are Riemann surfaces of infinite (both countable and continuum are exemplified) genus.

The following analogy is not entirely complete because we ignored the Frechet structure of the underlying manifold. Even in the sequence space case, there are multiple (countably many) coordinate patches that are necessary to cover the complement to the singularity locus. The above discussion allows us to single out the following case:

$$f : R \to V \tag{35}$$

where $R = R_\infty - \{z - n\}$, where the vector space $V$ admits a $C^*$- algebra connection, and $\pi_1$ is represented in $C^*$. Locally near the punctures $f(x) \sim (x - x_n)^{\alpha_{n,g}}$, $g \in \pi_1$.

## 6. Physical Interpretation

The flat bundle equations in this paper generalize perturbative quantum field theories into nonperturbative domains. Of particular importance is the construction of second quantized operators. We look at the energy momentum tensor

$$T_{\mu\nu}(p)[a, \frac{\delta}{\delta a}] = \sum \int (dq_1 \dots dq_n)(ds_1 \dots ds_m) \times \tag{36}$$

$$\times T_{\mu\nu}(p; q_1 \dots q_n; s_1 \dots s_m) \times \tag{37}$$

$$\times a(q_1) \dots a(q_n) \frac{\delta}{\delta a(s_1)} \dots \frac{\delta}{\delta a(s_m)} . \tag{38}$$

We observe that there is a version of multi-momentum configuration operators of the form

$$P_{\mu_1 \dots \mu_k}[a, \frac{\delta}{\delta a}] = \sum \int (dq_1 \dots dq_n)(ds_1 \dots ds_m) \times \tag{39}$$

$$\times P_{\mu_1 \dots \mu_s}(q_1 \dots q_n; s_1 \dots s_m) \times \tag{40}$$

$$\times a(q_1) \dots a(q_n) \frac{\delta}{\delta a(s_1)} \dots \frac{\delta}{\delta a(s_m)} . \tag{41}$$

This recovers the usual second quantized momentum, which is the starting point of the perturbative canonical quantization. Our notion goes beyond that. It allows us to discuss operadic spaces that result in particle collisions. The spectrum of momenta is defined as the spectrum of the second quantized momentum operator. In general, the spectrum of configurations of bound states is identified with the spectrum of the above multi-momenta operators. There is interplay with the operadic approach to the renormalization group [55]. It is part of the present formalism to incorporate operadic description of configuration spaces of particle configurations emerging out of a typical collision.

*Physical Amplitudes and Bases in the Space of Sections*

In the previous sections, we developed the theory of bundles of the form

$$\frac{\delta}{\delta f_z} \oplus F_{x_1,\ldots,x_n}[f] = \Omega_z \oplus F_{x_1,\ldots,x_n}[f] \tag{42}$$

where $f$ ranges over a function manifold, modeled locally on a function space $C^{\mathbb{L}}(M)$. Here we would like to state the following classification result. The basis elements of sections of such flat bundles are interpreted as amplitudes for $n \to m$ particle scattering. This is the nonperturbative definition of scattering theory applicable to theories with strongly bound particles in the initial and the final states. Below is a conceptual definition of a QFT in our present framework:

(1) Choose a function manifold $\mathbb{M}$.
(2) Choose a functional bundle on this manifold. This choice includes a choice in the tensor product $C(M) \otimes_\tau C(M)$ and a choice in the smoothness class $C^{\mathbb{L}}(M)$.
(3) Choose a flat connection on this bundle. This is the quantization step that defines the theory. This choice is highly constrained. In the examples above, we saw that it amounts to solving the flatness constraint at intersection points of components of singularity divisor of the connection.
(4) Find a basis for the solutions of the flatness equation. At this step we obtain scattering data for the strongly coupled theory.

## 7. Emergent Spacetime

It has been widely discussed that spacetime is an emergent notion, e.g., [1,56]. This discussion was mostly in the case of theories of gravity. This work suggests that conventional spacetime coordinates may be interpreted as effective labels emerging from spectral decompositions of functional evolution operators. Rather than being introduced a priori, spacetime-like structures could arise from the intrinsic geometry of functional bundles and their associated dynamical generators. The solutions of the evolution equations can sometimes be interpreted as happening in spacetime. Such events in particular encompass scattering events in QED, such as Compton, Moller and Bhabha scattering. Most importantly, it encompasses emission of photons in transition of bound states, such as the transition between hydrogen atom levels. This includes radiative corrections and Lamb shift [57]. In such emissions, the transition $F_{\alpha\beta}(x,y)[a,\ldots] \to F_{\alpha\beta\mu}(x,y,z)[a,\ldots]$ occurs. Here $a,\ldots$ stands for corresponding functional variables, while the transition happens between components of the bundle with two and three real particles.

The consistency conditions of our functional evolution equation can be formulated in a large variety of semigroups. The consistency conditions lead to the emergence of global time. This is the mechanism by which the global renormalization scale is set in multiscale events. If we have a process with multiple particles in the final state, it is a priori not set which scale one should use in the intermediate couplings in the process. We propose to identify the local time field with the global scale field of the process. If we consider the collision $p_1\, p_2$ of two particles with momenta $p_i$ and the space of all possible outcomes in this collision, then the flatness consistency condition sets the time scale for all these outcomes.

There is another version of evolution equation, namely,

$$i\frac{\delta}{\delta\tau_{\mu\nu}(x)}F = H_{\mu\nu}(x)F \tag{43}$$

where the local time $\tau$ is a rank 2 tensor, and the evolution operator is the second quantized energy momentum tensor $H_{\mu\nu}(x)[a, \frac{\delta}{\delta a}]$. This local time should be considered as a dual variable to dynamical energy momentum tensor $g_{\mu\nu}(x)$. It should be noted that we are still quantizing the theory in flat spacetime. In particular, this discussion perfectly applies to the ordinary QED as we know it in applications such as anomalous magnetic moment calculation [58] or Lamb shift calculation [57,59]. We propose to extend the quantization of QED in such a way that a dynamic scale field $g_{\mu\nu}(x)$ is added, and the global field that results from the consistency of the evolution equations is used to set the scale of renormalization in every part of the multiscale processes. In this sense, spacetime is emerging already in QED. It emerges as a result of evolution in functional space and then interpreting the results in terms of the usual momentum variables, as these are the quantities observed in experiment.

We would like to end this discussion with a toy model of the emergence of spacetime. In this toy model, the spacetime is two-dimensional, time is one-dimensional, and the energy momentum tensor is independent of functional time: $H_{\mu\nu}(q)[a, \frac{\delta}{\delta a}, g(q)] = H(q)[a, \frac{\delta}{\delta a}]$. Then we seek a solution as $F[a, T] = F[a]exp(i\int(dq)T(q)\lambda(q)$. For $\lambda(q)$ we get the equation

$$H(q)[a, \frac{\delta}{\delta a}]F[a] = \lambda(q)F[a]\,. \tag{44}$$

This is a functional eigenvalue equation for $\lambda(q)$. According to our discussion, this is a paradigmatic example of emergent manifold. In this case, it is just the graph of the function $\lambda(q)$.

The momenta or coordinates that are in use in calculations of scattering amplitudes are only indirectly related to the physically observable momenta, as measured in the detectors. These momenta comprise an internal manifold $M^3_{\vec{p}}$ from which the function space $\mathbb{M}$ is constructed. We only indirectly probe the properties of this internal manifold, $M^3_{\vec{p}}$, because its structure is related to the structure of fPDOs $\hat{\hat{P}}_\mu$, but it is not directly observable. In our picture of evolution in QFT the following elements are present:

(1) The internal manifold $M^3_{\vec{p}}$ is used to construct a function space $C^{\mathbb{L}}(M^3_{\vec{p}})$ and the functional manifold $\mathbb{M}$.
(2) On $\mathbb{M}$ a functional flat bundle is formulated. The bundle can have mild singularities along codimension 1 functional subvarieties.
(3) The bundle is classified according to a functional representation of $\pi_1(\mathbb{M})$. The bundle is essentially a topological, rigid object. It generalizes generating functional of perturbative QFTs and contains nonperturbative information about bound states.
(4) From $M^3_{\vec{p}}$ and $\mathbb{M}$ a set of second quantized fPDOs is constructed of the type $\hat{\hat{P}}_{\mu_1,\ldots,\mu_n}$. Once such operators are constructed, their evaluation on the sections of the flat bundle gives us tuples of finite dimensional spaces. These spaces are interpreted as configuration spaces of particles produced in the collision.

We showed in this paper that this generalization is consistent with the usual perturbative QED: with appropriate choice of $M^3_{\vec{p}}$ and $\mathbb{M}$, the above procedure gives the expressions studied in the literature. However, there are other choices that lead to nonperturbative parts in the properties of bound states.

The picture we described above is also consistent with perturbative QCD. Main predictions of perturbative QCD include RG evolution of partonic densities and spectra of heavy quarkonia. These properties are easily accounted for by low dimensional ansatzes for $\mathbb{M}$ and the bundle. The functional is in fact simple power law DGLAP ansatz in the perturbative regime, or the exponential solution for the BK equation, beyond perturbation theory.

*Physical Interpretation of the Notion of Emergent Spacetime*

The main physical claim of this paper is that evolution of particle systems takes place in an infinite dimensional functional space. It is only under very special circumstances it can be *interpreted* in terms of spacetime variables. These situations are the cases when functionals factorize into products of functionals that have simple behavior under application of variational multi-momentum operators $\hat{\hat{P}}_{\mu_1,\ldots,\mu_n}$. The fact that finite dimensional spaces can be embedded (preserving geometry) into infinite dimensional spaces is not by itself surprising. Our claim is that such interpretation is in fact necessary.

The physical phenomena that support our conclusion are in fact very well known in special cases. These include lineshape of hydrogen levels. The emission of photons from the excited hydrogen state can only be described in the combined function space of fermions and photons. In classical theory, this led to Wigner–Weisskopf lineshape theory [60] (for modern theory, see [61–67], full resurgent treatment of subexponential terms will be provided in further development of these ideas). In our language, this theory corresponds to consideration of the component $F_{\alpha\mu}(x,y)[f]$ of the flat bundle. But we need to systematically go beyond this truncation of the theory. Such generalization inevitably entails reconsideration of the notion of evolution in particle systems. This leads us to the following conclusion:

The manifold $M$ that underlies the construction of the functional manifold $\mathbb{M}$ is an auxiliary manifold. It is not directly related to physical spacetime, as observed in the experiments. It is rather an object whose structure we use in order to fix invariants of the functional flat bundles to the observable invariants, such as $\alpha(0)$. The physical spacetime *emerges* in the process of interpretation of solutions of the evolution equation according to the action of second quantized momenta operators $\hat{\hat{P}}_{\mu_1,\ldots,\mu_n}$.

Physically, the above conclusion entails lifting the actual lab experiment into the functional space. There are very vivid lab tours of modern precision experiments on hydrogen in Prof. Hansch's lab [68] (we note that similar spacetime interpretation of electron g $-$ 2 experiments in Gabrielse's lab [69,70] is more complicated due to the necessity to treat the electrons in the resonator consistently in second quantized formalism, by lifting them to the non-linear Fock space. The same applies to atomic interferometry experiments [71,72], which are crucial for $\alpha(\mu)$ measurements). According to our discussion, it is necessary to treat these experiments (individual absorption/emission events of laser photons and hydrogen atoms) as happening in an infinite dimensional function space. The geometry of the lab space is inherited from the geometry of the functional space. The line element in the functional space (the nonperturbative Fock space), in the case of pure gauge theory, has the form

$$ds^2 = \int DA_1 \int DA_2 \int_{x_1,\ldots;y_1,\ldots} G(x_1,\ldots,x_n;y_1,\ldots,y_m)[A_1,A_2]\times \tag{45}$$

$$\times\Delta F(x_1,\ldots,x_n)[A_1]\Delta F(y_1,\ldots,y_m)[A_2]. \tag{46}$$

In the presence of spinor fields, there are additional spinor variables in the above formula. $\Delta F(x_1,\ldots,x_n)[A]$ is a coordinate element, and $G(x_1,\ldots,x_n;y_1,\ldots,y_m)[A_1,A_2]$ is the metric functional on this non-linear manifold of states.

The mathematical necessity of such interpretation is demonstrated by the appearance of higher transseries terms in wave functionals $F(x_1,\ldots;y_1,\ldots)[\psi,A]$. These terms are a necessary ingredient of higher order terms in the exact treatment of lineshape theory. We obtain these terms only by considering exact evolution equations on functional manifolds. The effect of these terms is of course much more drastic in strongly coupled theories, which is of course the motivation of this work.

# 8. Comparison with Traditional Approaches

## *8.1. $G-2$ Computation*

Electron $g-2$ is extracted from the vertex function $\Gamma^{\mu}(p^2, p'^2, (p'-p)^2)$ at the point $m^2, m^2, 0$. In perturbation theory, $\Gamma^{\mu}(m^2, m^2, q^2)$ is decomposed into Master Integrals and is a component of a flat bundle that satisfies the following equation:

$$\frac{dI}{dq^2} = \sum \frac{A_n}{q^2 - r_n m^2} I \tag{47}$$

where $I$ is the vector of master integrals. The value $\Gamma^{\mu}(m^2, m^2, q^2 = 0)$ is then obtained as $Texp(\int_{\infty}^{0} \Omega(q^2) dq^2)$.

We show how to derive the usual expression for $g-2$ of electron from our new formalism. The starting point is the following functional:

$$F_0 = F(\vec{p})[a][T = -\infty] = u(\vec{p})a(\vec{p})|0> \tag{48}$$

where $u(\vec{p})$ is the usual plane wave solution to the Dirac equation. We take monodromy of our functional flat bundle around this point to arrive at

$$< F_0|Texp(\int d\tau \int d\vec{x} L_I(\tau, \vec{x}) \delta T(\tau, \vec{x})|F_0 > \ . \tag{49}$$

In this expression, there is ambiguity in the choice of contour of integration in the $\tau$ plane. We fix this using the flatness condition. The choice of $L_I(\tau, \vec{x})$ is such that it has a Stokes structure at $\tau = \infty$. We then can choose the contour of integration such that it goes along the real axis. In this way, we recover the expression for the bare fermion propagator, which can be used in combination with Ward identity to compute electron $g-2$ as done, e.g., in [8,73].

## *8.2. Lamb Shift Computation*

Now we add discussion of one- and two-loop Lamb shift calculation, which is traditionally performed along the lines of [4,57,59,74] or two-time perturbation theory. The comparison with this literature is more problematic. The issue is that the expressions in [4,57,59,74] are not gauge invariant. In the references [4,57,59,74] discussion of gauge invariance in the presence of 2-fermion wave functions is lacking. This precludes us from matching 1-1 the terms in our expansion with the terms in [4,57,59,74]. According to our formalism, one needs a more systematic treatment of symmetric functionals that contain higher vacuum polarization terms, e.g., of the form $|e^- e^+ \gamma >$, in order to compute higher terms in the Lamb shift. This is beyond the scope of the present article. The reason for this is that we did not include the treatment of the Local Renormalization Group, which is important for fixing the factorization ansatz for higher components in the wave functional $F(\vec{p}_1, \dots, \vec{p}_n)[a(\vec{p})]$, which is crucial at two loops and beyond. This issue will be addressed in future publications. At one loop, there are no such problems. One can start with the point $F_0(\vec{p})[a(\vec{p})] = \psi(\vec{p})a(\vec{p})|0>$ where $\psi(\vec{p})$ is the Dirac wave function, and then proceed as above, constructing the T-exponent and recovering the one-loop Bethe expression.

The usual starting expression

$$E_{SE} = (-ie^2) \int \frac{d^4k}{(2\pi)^4} < \psi_n|\gamma_\nu S(p+k)\gamma_\mu D_{\mu\nu}(k)|\psi_n > \tag{50}$$

where

$$\psi_n(\vec{p}) = \frac{P_n(\vec{p})}{(1 + a_n^2 \vec{p}^2)^{\gamma_n}} \tag{51}$$

for certain polynomial $P_n(\vec{p})$.

We give two derivations of the self energy at one loop—one in x-space, and one in p-space—as they have peculiarities. In order to give more details on the calculation of Lamb shift in our approach, we need to discuss the calculation of invariants of functional flat connections. Quantities

$$W(l) = Texp(\int_l \int d\vec{q}\Omega[T, a, \delta/\delta a]\delta T(\tau, \vec{q})) \quad (52)$$

and their matrix elements between states

$$F_{i,f} = F_X[a] \quad (53)$$

where $X$ stands for formal sum of spacetime variables. We will consider two alternative approaches in x- and p-spaces.

It is essential for our derivation that we integrate in the space of $T(r)$ from $T(r) = 0$ to $T(r) = 1$, i.e., between two points in the space $T(r)$ that correspond to constant functions. In general, we expect

$$T(q) = \sum \alpha^n T_n(q) \quad (54)$$

a perturbative expansion of $T(q)$. We are interested in monodromies around loops in the space $T(q)$. There is still certain freedom in our equations that has to do with the choice of space of local times $T(q)$. Our current understanding of this issue is that we should look at the spaces' logarithmic Puiseux series in $q$, which means that there is a set of algebraic varieties $L_i(q)$ such that near the zero-dimensional intersections the function $T(q)$ is representable by convergent series $\sum C(n_1, \ldots, n_d) \prod_i L_i^{a_i+n_i}(q)$ (where we regularized to remove possible logarithms). Then it is possible to realize the transition $T(q)$ from 0 to 1 in the whole space, as $\frac{1}{2\pi i} arg(log(H(q)))$ for $H(q)$ with logarithmic structures. We show how to use this to rederive one-loop corrections to Lamb shift.

As we know from finite dimensional theory, the moduli space depends only on the topological structure of the underlying manifold. We expect that the choice of the function manifold of local times is actually irrelevant for the calculation. It is rather the structure of the moduli space of flat functional bundles that matters for the structure of bound states. The full development of this theory requires a more detailed understanding of functional bundles and, in particular, the theory of tensor products of spaces of logarithmic Puiseux series on complex manifolds. These tensor products are usually non-unique and the full discussion will be given elsewhere. The logarithmic structure should be adapted to the logarithmic structure in the energy momentum tensor, i.e., in the expectation value $< \psi|\hat{T}_{\mu\nu}(x)|\psi >$. It is an elementary observation that the singularities of this function in p-space are logarithmic on a complement of a divisor. It is this logarithmic structure that plays a role in our construction of logarithmic Puiseux series. Logarithmic Puiseux series near complete intersection $L_a$ are series of the form $\sum_{(n_s)\in\mathbb{N}^k} f_{n_1,\ldots,n_k}(x_\parallel) \prod L_{a_s}^{\alpha_s+n_s}(x)$ where $x_\parallel$ are coordinates on the intersection $\cap L_{a_s}$. We regularized the possible logarithms $log^r(L_a(x))$ to power laws $L_a^\alpha(x)$ (dimensional regularization of logarithmic series).

There is one more choice that we have to discuss. Above we considered several versions of local time: $T(q)$, $t_{\mu\nu}(x)$, $\sigma_\mu(\tau, \vec{\xi})$. Below we will show that the choice $T(q)$ works at one loop. There is further choice $T_{\mu\nu}(q) = g_{\mu\nu}T(q)$ that allows us to reduce tensor time to scalar time. This choice is trivial in QED because $T_{\mu\nu}(x) = g_{\mu\nu}L_int(x)$. However, the choice matters in non-Abelian field theories.

8.2.1. X-Space Calculation

In x-space, $X = \cup(x_1, \ldots, x_n)$ and we are considering $F_X = \oplus F_{\alpha_1,\ldots,\alpha_n}(x_1, \ldots, x_n)[a, \delta/\delta a]$. We approximate

$$F_{\alpha_1}(x_1) \approx \psi_\alpha(x_1) \tag{55}$$

and neglect all higher components of the wave functional. We furthermore write

$$\Omega(q, p_1; r_1)[a, \delta/\delta a] = H^{QED}_{int}(t, q) \tag{56}$$

We will focus on identifying the terms that lead to the vertex correction diagram in the usual perturbative expansion. The integration in $T(q)$ is performed from $T(q) =$ to $T(q) = 1$.

$$W_{vert}(l) = \int H(t_1)dt_1 \int^{t_1} H(t_2)dt_2 \int^{t_2} H(t_3)dt_3 \tag{57}$$

The crucial aspect of our approach is the possibility of automatic choice of global time

$$T(x) \to t \tag{58}$$

in the processes. This is a consequence of integrability. For the Lamb shift, it means the following. We have wave functional

$$F[T] = \psi[T](\vec{x})a(\vec{x}) + \text{higher orders}\ . \tag{59}$$

The possibility of choice of global time allows us to transit to

$$F[T(\vec{x}) = t] = \psi(t, \vec{x})a(\vec{x}) + \text{higher orders} \tag{60}$$

which gives us the usual Dirac wave function.

8.2.2. P-Space Calculation

In this section, we show how we can reproduce, with certain assumptions, the usual perturbative expansion of the self energy

$$\Delta E_n =< \psi_n|\Sigma(p)|\psi_n >\ . \tag{61}$$

The self energy $\Sigma(p)$ comes from the double integral $\int_{-\infty}^{\infty} d\tau_1 \int_{-\infty}^{\tau_1} d\tau_2 \Omega[T(q) = \tau_1]\Omega[T(q) = \tau_2]$, where we chose $T(q) = \tau$ globally. Such choice is possible due to flatness.

The factors $< \psi_n|$ are coming from the asymptotic value of $F[T(q) \to \pm\infty]$. Essentially $F[T(q) \to -\infty] \to u(\vec{p})a(\vec{p})$. The reason for the appearance of hypergeometric functions $\psi_n$ in this expression is the choice of canonical quantization coefficients in $\hat{\vec{\psi}}(x)[a(\vec{p})]$. The flatness condition on the bundle does not by itself enforce this choice.

In our approach the following bundle on the space of $a(\vec{p})$ plays the central role

$$F_0[a] = \psi_n(\vec{p})a(\vec{p})\ . \tag{62}$$

Then we consider a path in the space of local times $T(q)$ that connects $T(q) = 0$ to $T(q) = 1$. We consider the evolution of $F_0$ under this flow

$$F[a][T] = Texp(\int d\tau \int d\vec{q}\Omega(q)\delta T(\tau, q))F_0[a]\ . \tag{63}$$

We then expand the T-exponent to 3rd order:

$$Texp(\int d\tau \int d\vec{q}\Omega(q)\delta T(\tau,q)) \approx \int_{-\infty}^{\infty} \Omega(\tau_1)d\tau_1 \int_{-\infty}^{\tau_1} \Omega(\tau_2)d\tau_2 \int_{-\infty}^{\tau_2} \Omega(\tau_3)d\tau_3. \quad (64)$$

The projection of $Texp(\int \Omega)F_0$ back to $F_0$ gives the usual self-energy and vertex correction at one-loop order. This calculation serves the purpose of rederiving, under the above assumptions, the known one-loop result [75].

According to our equations, time space field $T(q)$ emerges dynamically. This means that due to integrability we can choose $T(q) \to t$, a single global time variable for the interacting system, the hydrogen atom in this case. This gives us the solution of the above-mentioned potential ambiguity in the choice of RG scale in Bethe logarithm. The full discussion of fixing this RG scale is contained in a companion paper on Local Renormalization Group in QED.

*8.3. A Note on Comparison of Lamb Shift at Two Loops and Beyond*

The fundamental fact that hampers two-loop investigations of shifts of energy levels is that the proton radius correction is numerically larger than the two-loop effects, even the leading log ones. There are indications that the usual two-loop energy shift

$$\Delta E_n = < n|\Sigma^{(2)}(E_n)|n > \quad (65)$$

is problematic at two loops due to the issues with gauge invariance and the necessity to include higher component wave functions that have tensor structure like $\psi_{\alpha\mu}(x,y)$. But to test it experimentally, it is necessary to have a precise model of hadronic physics.

We suggest a framework for first principle investigation of this problem. It is not without its internal problems, however. For example, although the observables are rigid and sensitive only to topology of the manifold of local times, it still depends on the topology of this manifold, which is not known at present. This limits the predictive powers of this theory. In classical theory, everything is in principle computable, to any desired order in perturbation theory. This is not so in the theory discussed here. There are a few moving parts, like the topology of the manifold of local times, to which the leading observable numbers are insensitive, but which shows up at higher orders of perturbation theory.

Another somewhat puzzling prediction of our theory is that photons have subtle differences at two loops and beyond. In the usual theory, a photon is a well defined state, which is completely characterized by its momentum and polarization. In our theory, photons also have polarization and momentum. But additionally they have shape parameters that come from the non-uniqueness of solutions of the flat bundle equation

$$\frac{\delta}{\delta T(q)} \oplus F(x_1,\ldots,x_n;z_1,\ldots,z_m)[\bar{\eta},J,\eta] = \Omega(q) \oplus F(x_1,\ldots,x_n;z_1,\ldots,z_m)[\bar{\eta},J,\eta]\,. \quad (66)$$

We interpret these effects as being part of the linewidth structure of bound states. Precise resurgent analysis is necessary to resolve these issues. We note that experimentally the situation looks much better at the moment. The precision of experiments for 1s-2s and for 2s-4p transition exceeds the theory uncertainty budget.

## 9. Applications to Pure Gauge QCD

In this section, we consider applications of our equations to pure gauge QCD. In our functional quantization scheme the gauge fields become pseudodifferential operators of the type

$$G^a_\mu(x) = G^a_\mu(x)[a_s(\vec{p}), \frac{\delta}{\delta a_s(\vec{p})}] \quad (67)$$

that satisfy operator equations

$$[G^a_\mu(x), G^b_\nu(y)] = \delta^{a,b} D_{\mu\nu}(x, y) \ . \tag{68}$$

We are interested in the expectation values of operators between states. The main identification is as follows: collision events are eigenvalues of the functional monodromy operator, applied to the flat bundle. Collision events here include in particular stable bound states and decay channels for the unstable bound states. The computation procedure includes the following stages:

(1) Flat bundle is constructed, according to the flat connection (the generalized hamiltonian).
(2) Bases of sections of this flat bundle are constructed. The elements in the bases are indexed by tuples of manifolds, $M^n$, which are generalized spacetimes. The spacetimes are the emergent spacetimes discussed above. They correspond physically to the tuples of coordinates in the perturbative correlation functions, and generalize them to the nonperturbative domain. The issue is that these manifolds come with singular structure, and it is this singular structure (the ends of manifolds) that determines their geometry, topology, holomorphic structure, embeddings of the Euclidean spaces (the spaces of ordinary 3-momenta) and eventually the interpretation in terms of experimental data.
(3) Calculation of matrix elements between the basis states constructed in (2).
This includes the calculation of matrix elements of the energy momentum tensor

$$\hat{\hat{T}}_{\mu\nu}(x)[a, \frac{\delta}{\delta a}] = \sum \int^n \int^m T_{\mu\nu}(x; \vec{p}_1, \ldots, \vec{p}_n; \vec{q}_1, \ldots, \vec{q}_n) \times \tag{69}$$

$$\times a(\vec{p}_1) \ldots a(\vec{p}_n) \frac{\delta}{\delta a(\vec{q}_1)} \cdots \frac{\delta}{\delta a(\vec{q}_m)} \ . \tag{70}$$

The calculation of matrix elements proceeds as in the example below

$$< F|a(\vec{p}\frac{\delta}{\delta a(\vec{q})}|G> = \sum \int^n F_{n+1}(p, s_1, \ldots, s_n) G_{n+1}(q, s_1, \ldots, s_n)(ds) \ . \tag{71}$$

We would like to remark that this expression inevitably involves iterated integration of top forms on the manifold of momenta. We would like to interpret these expressions in terms of geometry of Banach spaces. The fact just mentioned suggests that the particular type of Banach spaces are of number theoretic nature because the integrals above can be interpreted as motives of open complex manifolds. In a sense, geometry of Banach spaces may inform the choice of the manifold of momenta $M^3_{\vec{p}}$, which the above formalism does not fix. We only note here that this formalism generalizes the usual calculations of the quarkonia in truncated QCD [76] and the approaches based on Bethe–Salpeter [77] and Schwinger–Dyson [78] hierarchies. In a sense, this is one of the points of this paper—there is a vast hierarchy of correlated wave functions that must be included in self consistent treatment of strong bound states. Our approach is most similar in spirit to [22].

## 10. Conclusions

In this paper, we analyzed examples of flat bundles on functional manifolds that generalize evolution equations of quantum field theories and that are valid in nonperturbative domain. We constructed examples for different situations with bundles of increasing complexity. In particular, we considered the case when the bundle is a direct sum of components with different numbers of spacetime variables. This corresponds to the physically relevant case of multicomponent wave functionals.

We discussed how the consistency of flat evolution equations leads to emergence of global time and renormalization scales in multiparticle processes. This discussion formalizes the emergence of spacetime in quantum field theories originally formulated in flat Minkowski spacetime. We think that this is of particular importance for QCD, because due to confinement phenomenon we can probe the phase space of partons only indirectly. In this theory, the internal space on which partons propagate can only indirectly be studied in scattering processes based on factorization properties of QCD. This space may have different properties, as the usual evolution equations [79] neglect many subleading terms. Our formalism allows for consistent treatment of evolution at arbitrary order. We should note that our conclusions are already valid and interesting in the case of QED of atoms and molecules. The difference with the standard theory [2] is of order $\alpha^2$ and thus requires precision studies in this case.

**Funding:** This research received no external funding.

**Data Availability Statement:** The original contributions presented in this study are included in the article. Further inquiries can be directed to the corresponding author.

**Conflicts of Interest:** The author declares no conflicts of interest.

## Appendix A. Manifolds Modeled on Sequence Spaces

In this section, we discuss the rigorous definition of flat bundles for sequence spaces.

**Definition A1** (manifold modeled on sequence spaces)**.** *Take a sequence Banach space $B$ where a point is denoted explicitly by $(x_\mu)$. A Banach manifold $\mathbb{M}$ is a topological space that has a countable covering by open sets $U_\alpha$ each isomorphic to the unit ball in $B$. Manifold structure is defined by the transition functions $\phi_{\alpha,\beta} : U_\alpha \cap U_\beta \to U_\alpha \cap U_\beta$.*

This definition in particular covers open manifolds, for example affine manifolds $\cap\{x : f_a(x) = 0\}$. The definition is only weakly dependent on the Banach space $B$. In fact, we can choose an incomparable Banach space $B'$ and use the ball of that space in order to construct the manifold. What matters to us is separability and the existence of countable cover. For example, take $l_p$ and $l_q$, $p < q$. It is well known that these spaces are incomparable, in the sense that the only operators between them are compact. But we can cover the unit ball of one space by a countable set of unit balls in the other. In this sense, we can replace the space in the definition, and use a different one. Differences start to appear if we try to use nonseparable spaces like $l_\infty$. To encompass this type of space, we would need to relax the countability requirement in our definition. Many of our constructions actually go through in this broader setting. One has to be careful with set theoretic issues as certain constructions may lead to statements that are independent of ZFC.

**Definition A2** (bundle on sequence manifold)**.** *Take a sequence manifold $\mathbb{M}$ as defined above. Take a sequence Banach space $C$, in general different from the base Banach space. A $C$-bundle on $\mathbb{M}$ is defined as a set of functions $f_\alpha : U_\alpha \to C$ that satisfy compatibility condition $f_\alpha(x) = f_\beta(\phi_{\alpha\beta}(x))$. If the space $C$ is a sequence space, we may write the trivialization $f_\alpha(x)$ explicitly as $f_i(x_\mu)$, where index $i$ refers to coordinates on $C$ and index $\mu$ refers to coordinates on $B$.*

**Definition A3** (flat bundle)**.** *Take a manifold $\mathbb{M}$ as defined above and a bundle with trivialization $f_i(x)$. The bundle is called flat if there exists an operator field $\Omega_{\mu,i,j}(x)$ such that*

$$\frac{\partial}{\partial x_\mu} f_i(x) = \Omega_{\mu,i,j} f_j(x) \tag{A1}$$

*subject to the compatibility condition*

$$\frac{\partial}{\partial x_\mu}\Omega_{\nu,i,j} - \frac{\partial}{\partial x_\nu}\Omega_{\nu,i,j} = \Omega_{\mu,i,k}\Omega_{\nu,k,j} - \Omega_{\nu,i,k}\Omega_{\mu,k,j} \tag{A2}$$

*For each* $\mu$*,*

$$\Omega_{\mu,i,j} : C \to C \tag{A3}$$

*so that the expression* $\Omega_{\mu,i,k}\Omega_{\nu,k,j}$ *is well defined.*

Of interest to us are the following flat bundles

$$\Omega_{\mu,i,j}(x) = \sum_{a=1}^{\omega} \frac{K_{\mu,i,j,a}(x)}{L_a(x)} \tag{A4}$$

where $L_a(x)$ are polynomials, while $K$ are entire.

Assume that we choose $A = a$ such that $\frac{\partial L_a}{\partial x_\mu}$ generate a space $B' \subset B$ such that $B = \mathbb{C} \oplus B'$ for all points $x \in \cap L_a$.

**Proposition A1.** *If the above assumption holds and* $L_a$ *are generic polynomials, then* $X_A = \cap_{x\in A}\{x : L_a(x) = 0]\}$ *is a Riemannian surface of continuum genus embedded smoothly in* $\mathbb{M}$*.*

**Proof.** By assumption, the set $X_A$ has charts that are isomorphic to a neighborhood in $\mathbb{C}$. We only need to prove that there is at most a continuum of them. But this is obvious because of our separability assumption on the manifold $\mathbb{M}$. □

**Remark A1.** *For generic* $L_a$*, the end of* $X_A$ *is accumulated by genus.*

We would like to prove a reconstruction theorem that allows us to identify $K_{\mu,i,j,a}(x)$ from its restrictions to points in the intersections $\cap_a L_a(x)$. Consider the intersections $Y_A = \cap_{a\in A} L_a(x)$ which are points. For generic $L_a$ and suitably chosen $A$, there is a continuum of such points. How this continuum is embedded in the separable space $\mathbb{M}$ is an open question. Full resolution inevitably entails the language of descriptive set theory (see [80], chapter 23). What we can say is that we have a continuum of operators $K_{\mu,i,j,a}(p), p \in Y_A$ and we would like to reconstruct the functions $K_{\mu,i,j,a}(x)$ from these data; while the full discussion of this problem is beyond the context, we would like to make the following observation.

**Proposition A2.** *If in the above reconstruction problem we are willing to consider nonseparable bundles, i.e., we allow the space* $C$ *be nonseparable, then the reconstruction problem has a formal solution.*

**Proof.** We can just identify each $K_{\mu,i,j,a}(p), p \in Y_A$ with a basis element in the (nonseparable) operator algebra and take formal closure of these elements. □

One of the motivations of this appendix is to establish DMPS formality theory [81] for Banach manifolds. We make the observation that the non-linear version of it, the non-Abelian cohomology of Simpson [82], does partially generalize to the Banach setting. Consider the holomorphic part of the Chern connection

$$\nabla = \partial_\mu + (h^{-1})^{a,\bar{c}}\partial_\mu h_{\bar{c},b} + \theta^a_{\mu,b} \tag{A5}$$

where $\theta$ is the Higgs field and $h^{a,\bar{b}}$ is the Hermitian metric.

**Proposition A3.** *The operator $\nabla$ is well defined on Banach manifolds if the inverse operator $h^{-1}$ is well defined.*

The conditions on $h_{a,\bar{b}}$ necessary for the existence of $(h^{-1})^{a,\bar{b}}$ depend on the base Banach space. Such inverse exists for example for Hereditary Indecomposable spaces, for which the operator has the form $h_{a,\bar{b}} = \lambda\delta_{a,\bar{b}} + k_{a,\bar{b}}$ where $k_{a,\bar{b}}$ is compact, see [83].

To obtain the solution of the flatness equation

$$\frac{\partial}{\partial z_{\bar{\nu}}}(h^{\bar{b},c}\frac{\partial}{\partial z_{\mu}}h_{a,\bar{b}}) + (\theta_{\mu})^c_b(\theta^{\dagger h}_{\bar{\nu}}) - (\theta^{\dagger h}_{\bar{\nu}})^c_b(\theta_{\mu})^b_a = 0 \tag{A6}$$

where $\theta^{\dagger h} = h^{-1}\theta^{\dagger}h,$ we consider the following heat flow

$$\frac{\partial h_t}{\partial t} = -2(F_h + [\theta, \theta^{\dagger h}])h_t\ . \tag{A7}$$

This equation is well defined in the Banach setting.

The key part of analysis of this flow is played by the polystability of the bundle. For this, we need to define $\mu(E) = deg(E)/rank(E)$. The degree is indeed well defined.

**Definition A4** (degree of a bundle)**.** *$deg(E)$ of a line bundle on a Banach manifold is the embedding type of its zero set into the manifold $\mathbb{M}$.*

Note that we cannot simply take the cardinality of this set. This set may have cardinality of continuum, with high Poincare rank, and have density points inside the separable manifold $\mathbb{M}$. We need to keep the geometry of this embedding. This definition is tightly related to the notion of Bourgain index [84].

In contrast, the notion of rank is not well defined. We link the problems of definition of rank to the well known issues in Banach homological algebra [85]. Another heuristic reason for the problem of this definition is the following. DGMS theory involves higher homotopy groups; while finite dimensional homotopy groups for Banach manifolds (or any other topological space) are well defined, this is not so for Banach homotopy groups, i.e., maps of unit spheres of different Banach spaces with incompatible norms (these sets are nullhomotopic, but not in a controlled way). Full discussion would require controlled homotopy theory and controlled homeomorphism theory of such spaces. Indeed, such a theory is developed by the author, but due to space and time limitations, we have to postpone its exposition to future publications. In general, we do not expect uniqueness of the solution of the flatness equation.

## Appendix B. Manifolds of Resurgent Functions

In this section, we build the beginnings of the theory of manifolds of resurgent functions. This example, while mathematically more complicated than the one in the previous section, is more close to the physics of bound states.

**Definition A5** (resurgent manifold)**.** *A manifold $M$ of (complex) dimension $n$ is said to be of resurgent type if it is a paracompact complex manifold of dimension $n$, and moreover in a neighborhood of each point $p \in M$ there exists a countable collection $N_a$ of resurgent manifolds of dimension $n-1$ such that the neighborhood of $p$ is isomorphic to a neighborhood of a point in the universal cover of $D^n - \cup N_a$, where $D^n$ is a polydisk. Moreover, in the case when $\cup N_a$ is dense in $D^n$, we require that for each finite distance $r$ from $p$ along $M$, there is only a finite set of $N_a$ that is visible. We will denote this manifold by $(M, N_a)$.*

**Remark A2.** *Note that for $n = 1$ we do not have to impose any special requirements on $N_a$, because these are just points in this case.*

**Definition A6** (regular resurgent function)**.** *Take a vector valued function $f_i(x)$ on a resurgent manifold $(M, N_a)$. We will call this function regular resurgent with index set $\alpha_a$ if for each component we have $f(x) = N_a(x)^{\alpha_a} R_1(x) + R_2(x)$, where $R_i(x)$ are holomorphic functions near $N_a$.*

**Definition A7** (manifold of resurgent functions)**.** *Let $K_i(x)$ be a set of resurgent functions. We define manifold of resurgent functions $\mathbb{M}$ to be the set of regular resurgent functions on $M$ that locally satisfy the condition $\int_{\zeta_i} K_i(x) f(x) d^n x = 0$, for some choice of contour $\zeta_i$.*

**Remark A3.** *Depending on values of the exponents $\alpha_a$ for $K_i$ and $f$, the contour $\zeta_i$ may or may not be allowed to end in the singularity $N_a$. This depends on the real parts to guarantee convergence.*

**Definition A8** (functional bundle on manifold of resurgent functions)**.** *Bundle on manifold of resurgent functions with structure group $C(x, y)$ of kernels that are resurgent on $M * M$ is a topological bundle with trivializations $F(x)[f]$ such that transition functions belong to $C(x, y)$.*

**Definition A9** (flat bundle on manifold of resurgent functions)**.** *Flat structure on a functional bundle is defined by the connection*

$$\nabla_x = \frac{\delta}{\delta f(x)} + \Omega(x; y, z)[f] \tag{A8}$$

*that satisfies flatness condition*

$$\frac{\delta}{\delta f(x)} \Omega(y; u, v) - \frac{\delta}{\delta f(y)} \Omega(x; u, v) = \tag{A9}$$

$$= \int_\zeta dw (\Omega(x; u, w) \Omega(y; w, v) - \Omega(y; u, w) \Omega(x; w, v)) \tag{A10}$$

Note that all the above definitions allow us to construct rather wild examples of nested resurgent manifolds. Wild in the sense of topology of the singularity set, the branching behavior is always combined power law. For physics applications we need a more restricted notion of singularity set.

**Definition A10.** *We call a set of codimension 1 manifolds $M_a(x_1, \ldots, x_n)$ elimination algebra $EA$, if it is closed with respect to elimination of variables in the degeneration conditions for intersections of $M_{a_s}(z_{s,1}, \ldots, z_{s,n})$ where $z_{s,i}$ is either $x_j$ or one of auxiliary variables $y_k$.*

**Theorem A1.** *Consider a functional flat bundle $F(x)[f]$. Construct the $EA$ generated from singularities of the defining functions $K_i$ for the manifold $\mathbb{M}$ and from the local power series expansions for $\Omega(z; x, y)$ (we assume here that $\Omega$ are regular resurgent). Then local solutions to the flatness conditions have singularities in this algebra $EA$.*

**Proof.** We sketch the idea of the proof of this theorem in the one-dimensional case $(M, N_a)$ where $N_a$ is a set of points on $M = \mathbb{CP}^1$. The manifold conditions are as follows:

$$\int_{\zeta_i} K_i(x) f(x) = 0 \tag{A11}$$

where $K_i(x)$ has singularities at a set of points $q_{i,a}$. The set of functions $f$ that satisfy these equations is the set of functions with regular singularities at both $N_a$ and $q_{i,b}$.

The flatness condition is as follows:

$$\frac{\delta}{\delta f(x)} F(u)[f] = - \int_v \Omega(x; u, v)[f] F(v)[f] \,. \quad \text{(A12)}$$

We seek a solution as series

$$F(x)[f] = \sum \int_{y_1,\dots} F_n(x; y_1, \dots, y_n) f(y_1) \dots f(y_n) \,. \quad \text{(A13)}$$

We see that the flatness condition is equivalent to an infinite hierarchy of coupled integral operators with kernels

$\Omega(x; u, v; y_1, \dots, y_n)$, where $\Omega_n$ are the kernels of expansion for $\Omega$.

The singularity sets on both sides of the bundle equation must coincide. On the right hand side, we have resurgent functions $F_n(x; y_1, \dots, y_n)$ convoluted with kernels $\Omega_m$ in the variable $v$. Convolution w.r.t. $v$ generates new singularities in the resulting function that are obtained as a result of elimination of variable $v$ from the singularity set of $\Omega_m$ and $F_m$. This is precisely the closure condition of the $EA$ algebra. Theorem follows. □

**Remark A4.** *From the proof we see that we can characterize the singularity set of the solution more precisely as a module over the $EA$ of singularities of $\Omega_m$.*

This theorem is of immediate physics interest. We obtain the following statement:

*Wave functionals of bound states are resurgent functional on the function manifolds of resurgent functions.*

This shows us that the familiar perturbative expansions give us but a few layers in these infinite towers of resurgent expansion functionals, truncated according to the power counting in coupling constant. In contrast, for strongly interacting bound states, such as the proton, we need to take into account the whole tower of resurgent layers.

**Remark A5.** *Note that in our definition of $EA$ we did not need algebraicity of singularity components $L_a$. Our theorem is valid even in the case of "non well founded" algebras where there are no algebraic generators from which the $EA$ is constructed by closure. We hypothesise that this is in fact happening in physics, i.e., the physical solutions are built on resurgent manifolds with transcendental singularity components. The familiar Landau polynomials are merely low order truncations of these transcendental objects.*